\documentclass[referee]{raa}            

\usepackage{graphicx,times}             
\usepackage{natbib}
\usepackage{amssymb,amsmath}
\bibpunct{(}{)}{;}{a}{}{,}

\usepackage[a4paper=true,dvipdfm=true,pagebackref=true]{hyperref}
\hypersetup{colorlinks = true, linkcolor = green, anchorcolor = red, citecolor = blue, filecolor = red, pagecolor = red, urlcolor = red}

\begin{document}

   \title{Calibration of X-ray telescope prototypes at PANTER
}

   \volnopage{Vol.0 (20xx) No.0, 000--000}      
   \setcounter{page}{1}          

   \author{Ying-Yu Liao
      \inst{1,2}
   \and Zheng-Xiang Shen
      \inst{1,2}
   \and Jun Yu
      \inst{1,2}
   \and Qiu-Shi Huang
      \inst{1,2}
   \and Bin Ma
      \inst{1,2}
   \and Zhong Zhang
      \inst{1,2}
   \and Xiao-Qiang Wang
      \inst{1,2}
   \and  Kun Wang
      \inst{3}
   \and  Chun Xie
      \inst{4}
   \and  Vadim Burwitz
      \inst{5}
   \and  Gisela Hartner
      \inst{5}
   \and  Marlis-Madeleine La Caria
      \inst{5}
   \and  Carlo Pelliciari
      \inst{5}
   \and  Zhan-Shan Wang
      \inst{1,2}
   }

   \institute{Key Laboratory of Advanced Micro-Structured Materials, Ministry of Education, Tongji University, 200092 Shanghai, China; {\it wangzs@tongji.edu.cn}\\
        \and
             Institute of Precision Optical Engineering, School of Physics Science and Engineering, Tongji University, 200092 Shanghai, China\\
        \and
             School of Mechanical Engineering, Tongji University, 200092 Shanghai, China\\
        \and
             Sino-German College of Applied Sciences, Tongji University, 200092, Shanghai, China\\
        \and
             Max-Planck-Institute for extraterrestrial Physics, Giessenbachstr., 85748 Garching, Germany\\
\vs\no
   {\small Received~~20xx month day; accepted~~20xx~~month day}}

\abstract{ We report a ground X-ray calibration of two X-ray telescope prototypes at the PANTER X-ray Test Facility, of the Max-Planck-Institute for Extraterrestrial Physics, in Neuried, Germany. The X-ray telescope prototypes were developed by the Institute of Precision Optical Engineering (IPOE) of Tongji University, in a conical Wolter-I configuration, using thermal glass slumping technology. Prototype \#1 with 3 layers and Prototype \#2 with 21 layers were tested to assess the prototypes’ on-axis imaging performance. The measurement of Prototype \#1 indicates a Half Power Diameter (HPD) of 82" at 1.49 keV. As for Prototype \#2, we performed more comprehensive measurements of on-axis angular resolution and effective area at several energies ranging from 0.5-10 keV. The HPD and effective area are 111" and 39 $cm^{2}$ at 1.49 keV, respectively, at which energy the on-axis performance of the prototypes is our greatest concern.
}

   \authorrunning{Y.-Y. Liao, et al. }            
   \titlerunning{Calibration of X-ray telescope prototypes}  

   \maketitle

%
%
\section{Introduction}           
\label{sect:intro}

In grazing incidence X-ray observations, imaging X-ray telescopes (IXT) employing Wolter-I configuration and its optimization solutions have been developed for half century. The Wolter-I configuration, consists of a pair of co-axial and con-focal paraboloid and hyperboloid mirrors, was proposed by Wolter in 1952 (\citealt{Wolter V. H.+1952}). X-ray telescopes using focusing grazing incidence optics, like Wolter-I configuration, were noted by Giacconi and Rossi in 1960 (\citealt{Giacconi R.+etal+1960}). To obtain large collecting area, a multilayer nested Wolter-I configuration was described by Van Speybroeck and Chase in 1972 (\citealt{VanSpeybroeck L. P.+etal+1972}). To improve the angular resolution of the X-ray telescope, many optimization solutions were proposed, amongst them the Wolter-Schwarzschild geometry (\citealt{Wolter V. H.+1952}), polynomial geometry (\citealt{Werner W.+etal+1977, Burrows C. J.+etal+1992, Conconi P.+etal+2002}), double hyperboloid geometry (\citealt{Thompson P. L.+etal+1999, Harvey J. E.+etal+2001}) and Modified Wolter-Schwarzschild geometry (\citealt{Saha T. T.+etal+2014}). To reduce the difficulty and cost of mirror fabrication, the conical Wolter-I configuration was put forward in the 1980s (\citealt{Petre R.+etal+1985, Serlemitsos P. J.+1988}). In recent years, two optimization solutions were proposed by the Institute of Precision Optical Engineering (IPOE), which are a hybrid and a sectioned configuration (\citealt{Chen S.-H.+etal+2016, Liao Y.-Y.+etal+2019}). The hybrid configuration uses one conical surface and one quadratic surface, while the sectioned configuration is based on a conical Wolter-I configuration with sectioned secondary mirrors. China has made large step forward in the field of non-imaging X-ray telescopes, which is the well-known Hard X-ray Modulation Telescope (Insight-HXMT) (\citealt{Li T.-P.+etal+2017}). However, until now China has been involved in international cooperation to fabricate imaging X-ray telescope in Wolter-I configuration. In the past decade, several IXT missions have been proposed, in which China is involved. The X-ray Timing and Polarization (XTP) (\citealt{Dong Y.+2014}) that was transformed into the enhanced X-ray Timing and Polarimetry (eXTP) (\citealt{Zhang S.-N.+etal+2016}), now in Phase B, is a mission that has been selected as a successor of the Insight-HXMT. A different mission Einstein Probe (EP) (\citealt{Yuan W.-M.+etal+2015}), also in Phase B, is designed to discover transients and monitor variable objects in the 0.5-4 keV X-ray band, at a sensitivity higher by one order of magnitude than those of missions currently in orbit. The Hot Universe Baryon Survey (HUBS) mission (http://heat.tsinghua.edu.cn/~hubs/) is being proposed to primarily address the issue of “missing baryons” in the local universe.

%
   \begin{figure}[htbp]
   \centering
   \includegraphics[width=12 cm, angle=0]{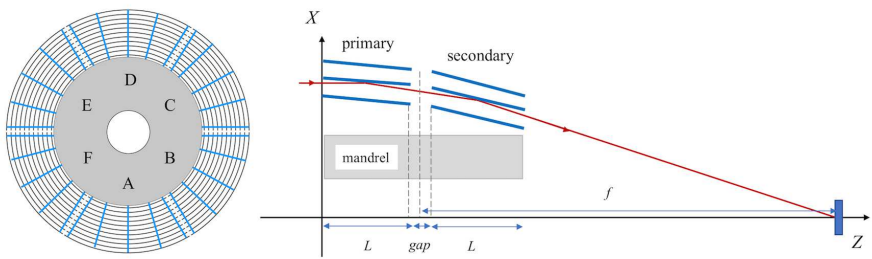}
   \caption{Schematic of the prototype, the entrance aperture (left) and the cross-section profile (right). }
   \label{Fig1}
   \end{figure}

At the IPOE of Tongji University, we have been developing imaging X-ray telescopes independently for over a decade (\citealt{Wang Z.-S.+etal+2014, Shen Z.-X.+etal+2018}). The thermal slumping technology is utilized to fabricate mirror substrates, which was firstly proposed in an experimental KB telescope for the EUV and soft X-ray bands (\citealt{Labov S. E.+1988}), and developed for the HEFT and NuSTAR optics (\citealt{Craig W. W.+etal+2011, Koglin J. E.+etal+2004}). Two slumped glass prototype optics modules that we fabricated were tested at the PANTER X-ray Test Facility, Prototype \#1 with 3 mirror layers and Prototype \#2 with 21. Both of them use the conical Wolter-I configuration, as illustrated in Fig. 1. The confocal and concentric layers share the common focal length f, which is defined as the axial distance between the focus to the midpoint (principal plane) between the primary and secondary mirrors. The mirrors are nested tightly to maximize the on-axis collecting area. The prototype mirror module consists of six sectors, Sector A-F, each of which uses five graphite spacers to stack the mirrors from the mandrel shell by shell. In Table 1 the characteristics of these two prototypes are summarized.

Prototype \#1 has a better HPD benefiting from smaller diameter. Prototype \#2 with more nested mirrors was tested more comprehensively to assess the on-axis imaging performance, thus acquiring reliable feedbacks to improve the fabrication of the imaging X-ray telescope.


%
\begin{table}[htbp]
\begin{center}
\caption[]{ Characteristics of Prototype \#1 and \#2.}\label{Tab:publ-works}


 \begin{tabular}{clcl}
  \hline\noalign{\smallskip}
   &  Prototype \#1      & Prototype \#2                    \\
  \hline\noalign{\smallskip}
Number of layers N
  & 3 & 21      \\ 
Focal length f (mm)
  & 2052.5 &   2052.5 \\
Diameter D (mm)
  & 104-109 &   104-150 \\
Mirror length L (mm)
  & 100 &   100 \\
Mirror thickness t (mm)
  & 0.3 &   0.3 \\
Mirror coating
  & Pt &   C/Ni/Pt \\
Grazing angle $\alpha$ (deg)
  & 0.365-0.379 &   0.365-0.522 \\
HPD (")
  & 82 at 1.49 keV &   111 at 1.49 keV \\
Effective area ($cm^{2}$)
  & - &   39 at 1.49 keV \\
  \noalign{\smallskip}\hline
\end{tabular}
\end{center}
\end{table}

\section{MEASUREMENT SETUP}
\label{sect:Obs}

%
   \begin{figure}[bp]
   \centering
   \includegraphics[width=12 cm, angle=0]{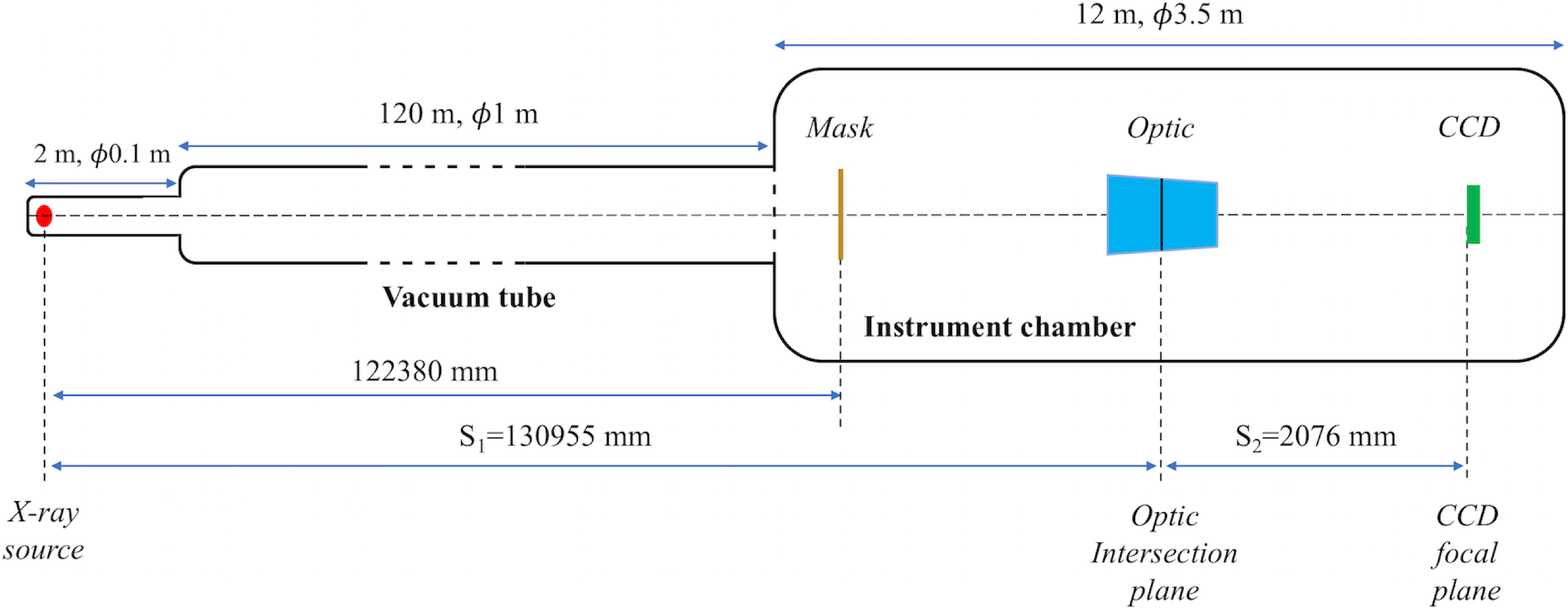}
   \caption{ Schematic of the PANTER X-ray Test Facility, where the quasi-parallel X-ray beam is achieved. }
   \label{Fig2}
   \end{figure}

The PANTER X-ray Test Facility (\citealt{Freyberg M. J.+etal+2005, Freyberg M. J.+etal+2008}) was built to develop and characterize the ROSAT optics. It is a laboratory of the Max-Planck-Institute for Extraterrestrial Physics (MPE). PANTER has been utilized successfully for developing and calibrating X-ray astronomical instrumentation for observatories such as EXOSAT, Chandra (LETG), BeppoSAX, XMM-Newton, Swift (XRT), eROSITA, etc., except for ROSAT. PANTER has a beam path length of 123.6 m, thereby providing a wide aperture quasi-parallel X-ray beam. This long beam length is realized by utilizing a vacuum tube (length of 120 m and diameter of 1 m) between the X-ray source and the instrument chamber (length of 12 m and diameter of 3.5 m). The instruments in the chamber can be translated and rotated by means of manipulators driven by stepper motors with a typical accuracy of $<$3 µm. In the tube and chamber the vacuum degree can be kept at a pressure of ${<10^6}$ mbar during measurement. The schematic of the PANTER X-ray Test Facility is shown in Fig. 2.

A dedicated backside illuminated PN-CCD camera is utilized, called TRoPIC (Third Roentgen Photon Imaging Counter) (\citealt{Burwitz V.+etal+2013}), specially developed for calibration measurements for eROSITA. TRoPIC has a pixel size of 75 $\mu$m and an array of 256 $\times $ 256 pixels, giving it a field of view of 19.2 mm $\times $ 19.2 mm. In front of the prototypes is a movable mask installed, determining which sector of the optic is illuminated. 
For both prototypes, we intentionally fabricated one of the six sectors using mirrors with the best quality and another one with the worst quality. As for the figure error of mirrors, we utilized the linear variable differential transformer (LVDT) to make dense azimuthal and axial scans of the mirror segment surface height profile during Prototype assembly. As a result, Sector A' of Prototype \#1 has the best quality. For Prototype \#2, Likewise, Sector A, B and C are characterized by the best, the worst and the moderate quality, respectively. To study both global and local performance of the two prototypes, the four sectors mentioned above are tested specifically except for the full aperture of prototypes. The PSF of Prototype \#1 was measured at 1.49 keV and 8.04 keV. More comprehensively, the PSF and effective area of Prototype \#2 were determined at several energies, 525 eV, 1.49 keV, 4.51 keV, 4.9 keV, 8.04 keV and 9.9 keV. 

%
   \begin{figure}[htbp]
   \centering
   \includegraphics[width=10 cm, angle=0]{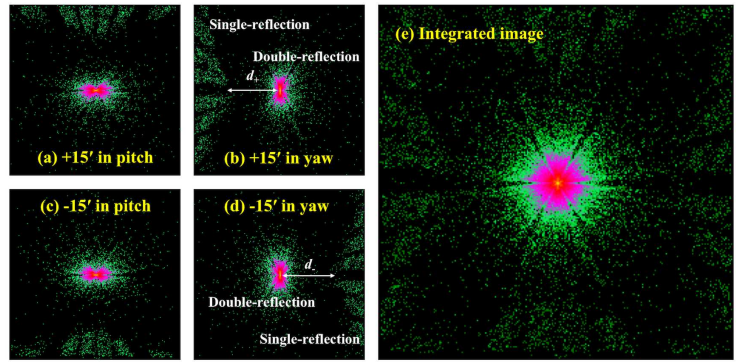}
   \caption{ Images of Prototype \#2 by the Burkert test. Fig. 3a, 3b, 3c, and 3d are the individual images and Fig. 3e is the integrated image. }
   \label{Fig3}
   \end{figure}

As a prerequisite of a reliable measurement, X-ray alignment of Prototype and the optical axis should be performed carefully. The Burkert test (\citealt{Menz B.+etal+2013}) is an efficient alignment method at the PANTER, by utilizing the different behaviors of the single-reflection and double-reflection with respect to the off-axis angle, with the optic changing its attitude in pitch and yaw direction. The double-reflection rays contribute to the image on the focal plane, while the single-reflection is reflected only once either by the primary mirror or the secondary mirror. The position of the image depends on the chief ray, which is determined by the source and the center of the optic. For a fixed source, the chief ray does not change as the attitude of the optic changes. As a result, the centroid of the image is also unchanged on the focal plane ideally, even though its shape changes. In other words, the position of the normal image on the focal plane is in general independent of the off-axis angle. While the position of the single-reflection is sensitively dependent on the off-axis angle. Therefore, the distance between the single-reflection and double-reflection can be significant at a large off-axis angle. In addition, in the case of a perfect alignment, the distances should be identical at a pair of opposite off-axis angles because of symmetry. We define the distance between the single-reflection and double-reflection as d+ in the case of off-axis angle of $\theta$, and then rotate the optic in the opposite direction to acquire d- at -$\theta$ off-axis angle. In an iterative process, we adjust the attitude of the optic until the equation d+=d- makes sense. This process will provide $<$1' knowledge of the pitch and yaw of the optic, which is far smaller than the grazing incident angle of the optic. Prototype \#2 was finely aligned by the Burkert test. In Fig. 3 is the images of Prototype \#2 at opposite off-axis angles in pitch and yaw directions by Burkert test. The images were taken when the optic was set at four equally sized and opposite off-axis angles $\theta$, which were ±15' in yaw and pitch direction, respectively.

\section{MEASUREMENT RESULTS}
\label{sect:data}

\subsection{Focus Search of Prototype \#2}
As shown in Fig. 2, with a ﬁnite source distance S1 the image (optic-focal plane) distance S2 will be slightly longer than the nominal focal length f, as given by the thin lens equation. 
\begin{equation}
  \frac{1}{S_{1}}+\frac{1}{S_{2}}=\frac{1}{f}
\label{eq:LebsequeI}
\end{equation}

%
   \begin{figure}[hbp]
   \centering
   \includegraphics[width=10 cm, angle=0]{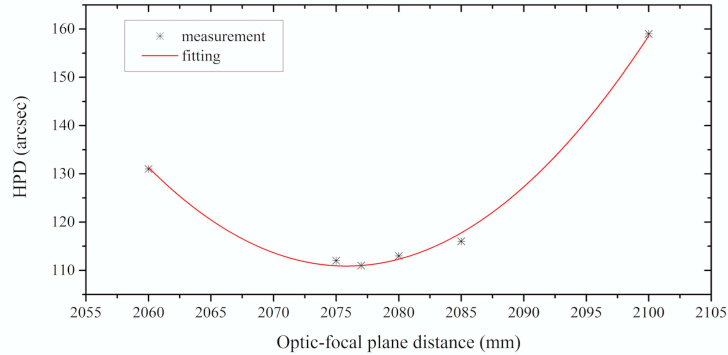}
   \caption{ Variation of HPD of Prototype \#2 with optic-focal plane distance at 1.49 keV, indicating the best focus at a optic-focal plane distance of 2076 mm. }
   \label{Fig4}
   \end{figure}

A focus searching process was performed at 1.49 keV, by adjusting the optic-focal plane distance to determine the minimum HPD as shown in Fig. 4. The theoretical image distance is 2085 mm, while the measured image distance was slightly shorter, giving it a value of 2076 mm according to the best fit to the focus curve. The minimum HPD of 111" at 1.49 keV is determined at the best focal-plane distance of 2076 mm, indicating the measured EEF in Fig. 5, with the simulated EEF for comparison. The preliminary assessment by means of a ray-tracing program predicted an HPD of 101" of the prototype. In addition to the 3.9' divergence of the X-ray beam, the figure error of every individual mirror (30"-180") and the residual 30" misalignment after alignment process were also taken into account in the simulation. The 3.9' beam divergence is the angular diameter of the incident beam (150 mm in diameter) with respect to the source-optic distance (130955 mm). The figure error of the mirrors was evaluated by a Linear Variable Differential Transformer (LVDT), which is an in-situ measurement system utilized to measure the mounted mirrors during assembly process (\citealt{Koglin J. E.+etal+2011}).The value of 30"-180" are used to assess the figure error of each mirror, which are the predicted results by ray-tracing program. These results are based on the combination of one mirror with measured figure error, and another mirror in a perfect conical approximation geometry.

%
   \begin{figure}[htbp]
   \centering
   \includegraphics[width=9 cm, angle=0]{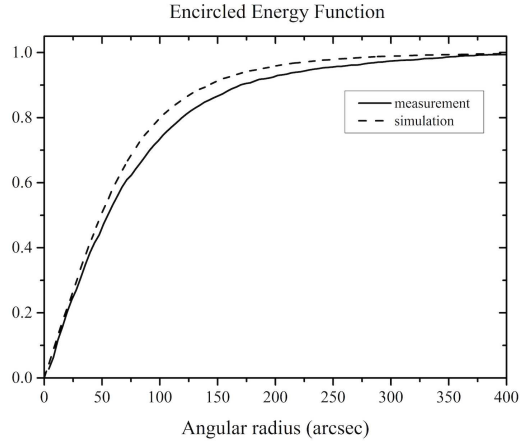}
   \caption{ Simulated and measured EEFs of Prototype \#2, indicating a simulated HPD of 101" and a measured HPD of 111".}
   \label{Fig5}
   \end{figure}

The deviation of the image distance can be attributed to a combination of imperfect mirrors and small errors in the mirror assembly process. These are the issues that will be improved in the future fabrication processes. As a result of the longitudinal deformation of the mirrors and the assembly errors, the kink angle (twice the grazing angle theoretically) between the primary and secondary mirrors could vary. The deviation could change the image distance but would not change the PSF significantly, because the small deviation is negligible compared with the grazing angle. Approximately, 28" deviation of the kink angle could introduce a 9-mm deviation of the image distance, estimated by Eq. (2), where $\alpha$ is the grazing angle, $\beta$ is the deviation of the kink angle, and f is the focal length.

\begin{equation}
  \Delta =f\cdot (1-\frac{tan(4\alpha )}{tan(4\alpha +2\beta)})
\label{eq:LebsequeI}
\end{equation}

\subsection{Out-of-focus rings of Prototype \#2}
Measurements of the out-of-focus rings (\citealt{Misaki K.+etal+2008}) were performed by moving the detector, to positions of ±150 mm and ±120 mm from the best focus. In Fig. 6 the out-of-focus rings of Prototype \#2 and the azimuthal intensity distribution therein are clearly visible, also the shadows of the support structure and graphite spacers (the entrance aperture of Prototype \#2 is illustrated in Fig. 1) are visible.

%
   \begin{figure}[htbp]
   \centering
   \includegraphics[width=12 cm, angle=0]{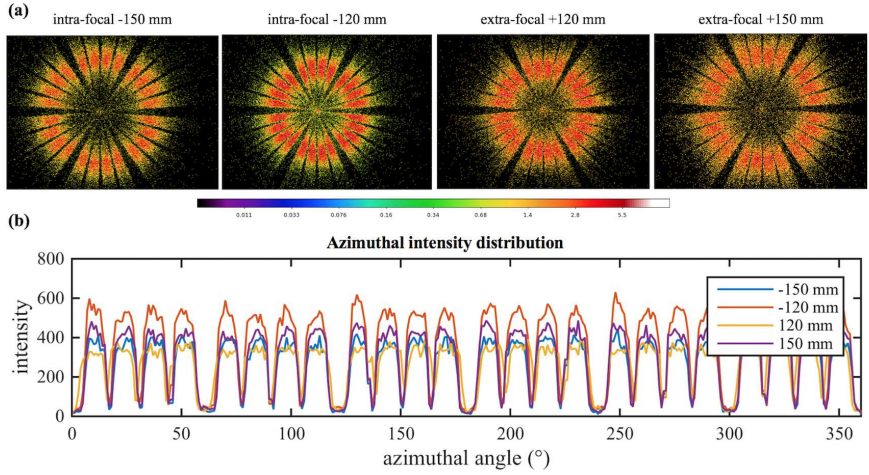}
   \caption{ (a) Out-of-focus rings of Prototype \#2, intra-focal (-120mm, -150 mm) and extra-focal (+120mm, +150 mm). (b) azimuthal intensity distribution of the out-of-focus rings.}
   \label{Fig6}
   \end{figure}

By means of out-of-focal test, the rings can be analyzed quantitatively with radial proﬁles to assess the local performance of the optic, which means the assessment of the local performance can be made by full flood illumination rather than by pencil beam. The rings are also utilized to determine the effective area, thus avoiding detector pile-up effects. The effective area was measured at 525 eV, 1.49 keV, 4.51 keV, 4.9 keV and 8.04 keV, compared with results from simulation (see Fig. 7). The effective area is derived using Eq. (3), where EA represents effective area, A and B are the photon counting rate with and without optic, respectively. S is the geometric area of the detector CCD, and C is the correction factor determined by the time-stability of the X-ray beam. The time-stability was acquired by measuring the direct beam before and after the effective area test of each energy using TRoPIC. The intensity of the beam is considered as uniform, because the X-ray beam at PANTER is of good uniformity. As shown in Fig. 8, the uniformity test of the X-ray beam was performed by measuring the direct beam using TRoPIC. Each square is corresponding to one field of view of TRoPIC.

\begin{equation}
  EA=\frac{A}{B}\times S \times C
\end{equation}

The measured effective area at 1.49 keV is 39 $cm^{2}$, our reference energy for the on-axis performance of the prototypes. This value is about 5$\%$ lower than expected, which is mainly ascribed to the epoxy glue blocking the light path. The epoxy glue is used to bond the mirrors and graphite spacers, a few (0.2-0.3 mm in width visibly) of which spilled out and contaminated the mirrors during epoxy pasting and curing process. In this case, the excess epoxy glue can degrade the effective area. As shown in Fig. 6, the out-of-focus rings indicate a lower intensity of the area close to the graphite spacers.

%
   \begin{figure}[htbp]
   \centering
   \includegraphics[width=9 cm, angle=0]{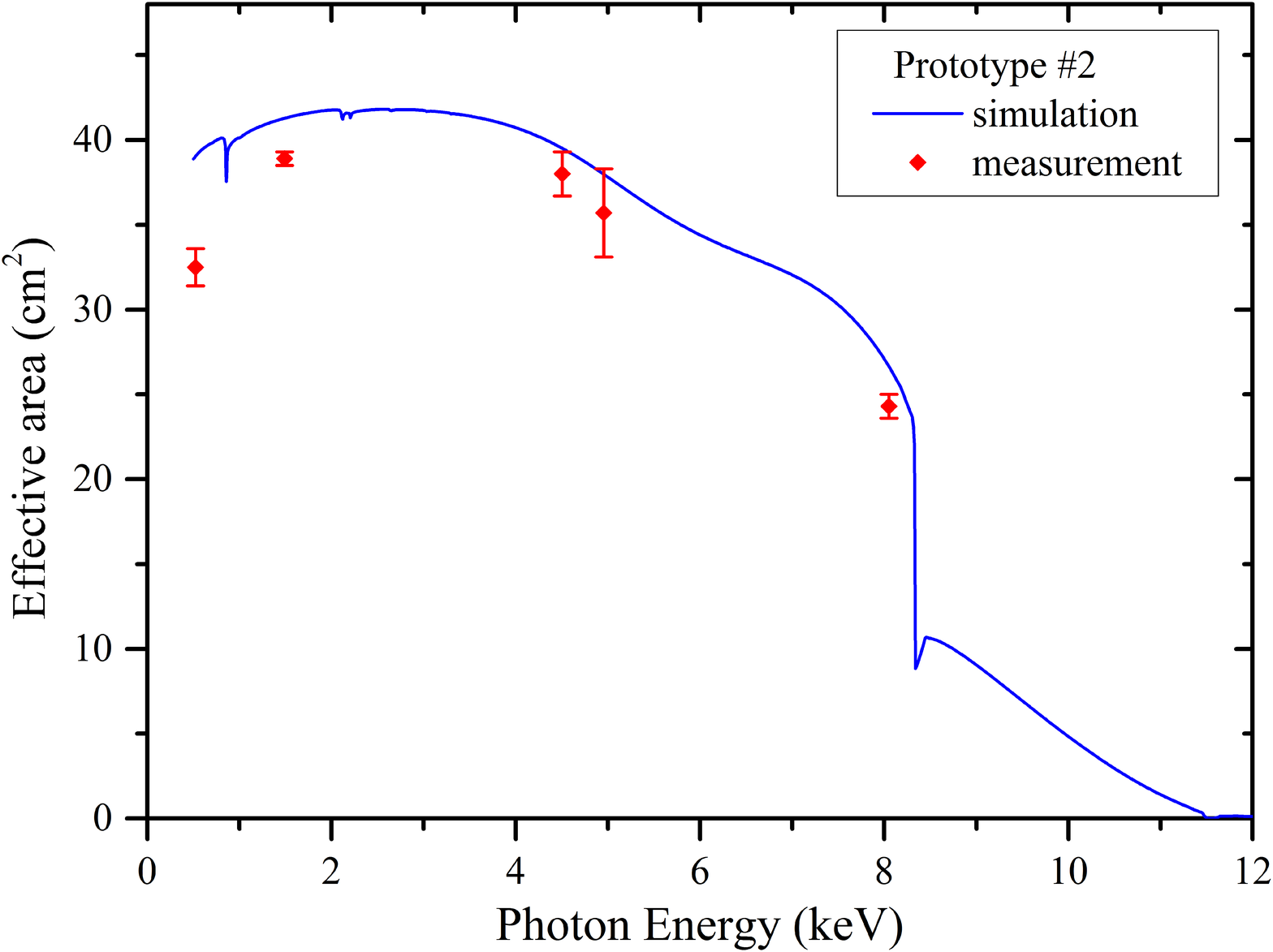}
   \caption{  Comparison of simulated and measured effective area of Prototype \#2. The measured effective area at 1.49 keV is deviated from expectation by 5 $\%$ as a result of epoxy glue.}
   \label{Fig7}
   \end{figure}
   \begin{figure}[htbp]
   \centering
   \includegraphics[width=10 cm, angle=0]{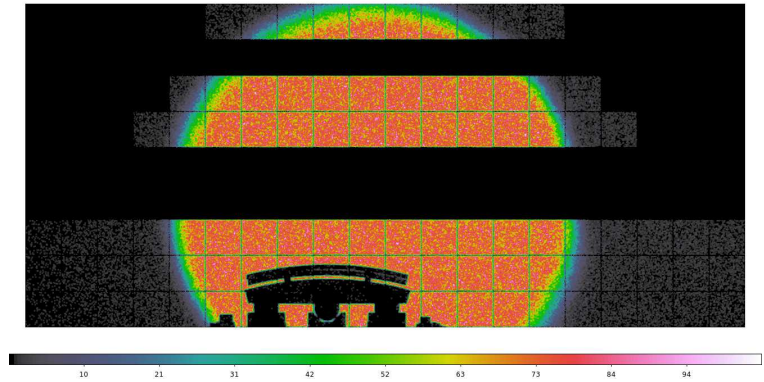}
   \caption{  Uniformity test result of the X-ray beam using TRoPIC (credit: Gisela Hartner, the PANTER X-ray Test Facility).}
   \label{Fig8}
   \end{figure}
%

%

   \begin{figure}[hbp]
   \centering
   \includegraphics[width=9 cm, angle=0]{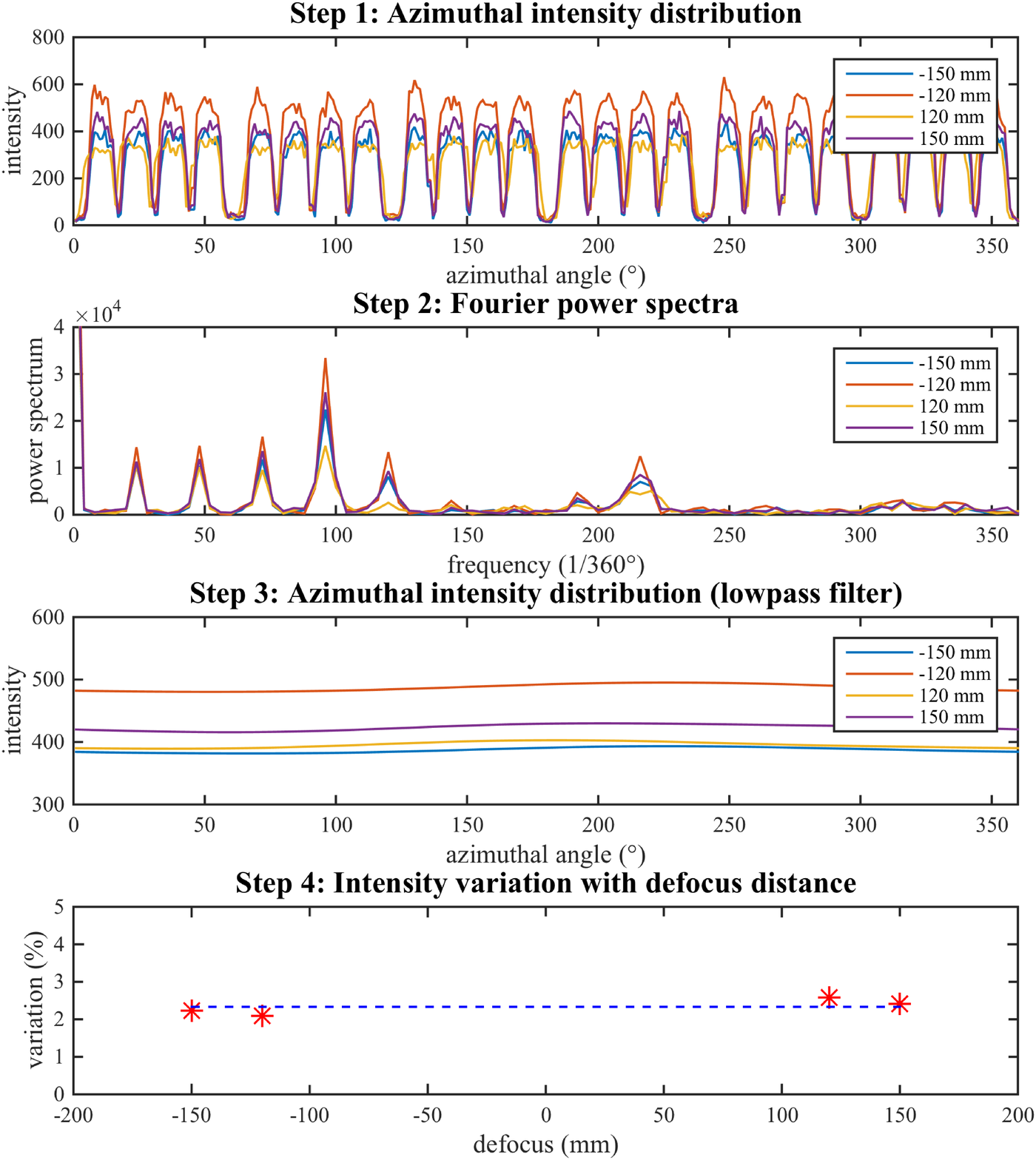}
   \caption{ Analysis of the azimuthal intensity of the out-of-focus rings, indicating minor changes of $<$3 $\%$ in the azimuthal intensity distribution.}
   \label{Fig9}
   \end{figure}

Apart from the Burkert test, there are other two methods for X-ray alignment methods that are used at the PANTER, the Cross-scan and the Egger-Menz tests (\citealt{Menz B.+etal+2013}). The Cross-scan is based on a symmetry of the image blurring behavior for increasing off-axis angles, which is measured by the HPD of the point image. The Egger-Menz test symmetrizes the azimuthal intensity distribution that yields in symmetric effective areas. Compared with the Burkert test, the Cross-scan and the Egger-Menz test are more precise but time-consuming. Nevertheless, the Egger-Menz method can be utilized to assess the alignment of the mirrors based on the out-of-focus rings. The Egger-Menz test is designed as a fine alignment process using the symmetry of effective areas. In another word, for a perfectly aligned mirror the azimuthal intensity distribution of the effective area is homogeneous, while for an off-axis aligned mirror the intensity distribution becomes elliptical. According to the out-of-focus rings, the azimuthal intensity is integrated, thus acquiring the azimuthal intensity distribution in Fig. 6b. Learning from the Egger-Menz method, the azimuthal intensity distribution is analyzed based on the Fast Fourier Transform, for which the method is shown in Fig. 9. By using a low-pass filter, the shadows of the support structure and graphite spacers are removed. After that, the intensity distribution shows only minor variation of less than 3 $\%$ in azimuth, which indicates the mirrors are well-aligned.

\subsection{Measurement of Prototype \#1}
Likewise, the smallest PSF of Prototype \#1 was found at an image distance of 2080 mm. The simulated and measured EEFs of Sector A' and full aperture of Prototype \#1 are shown in Fig. 10.

%
   \begin{figure}[htbp]
   \centering
   \includegraphics[width=12 cm, angle=0]{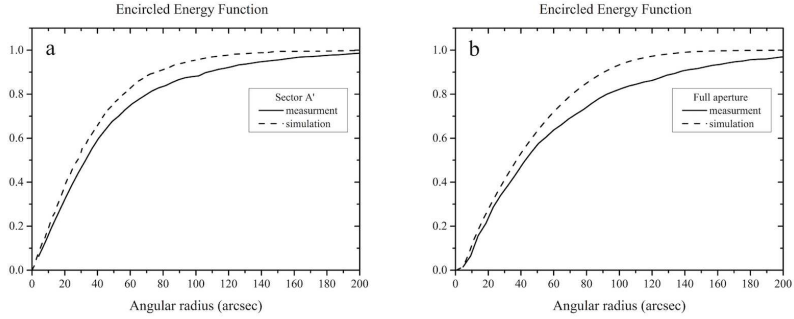}
   \caption{ (a) Simulated and measured EEFs of Sector A’ of Prototype \#1, indicating an HPD of 56" and 67" at 1.49 keV, respectively. (b) Simulated and measured EEFs of the full aperture of Prototype \#1, indicating an HPD of 77" and 82" at 1.49 keV, respectively.}
   \label{Fig10}
   \end{figure}

\subsection{Overview of the Measurement Results}

%
\begin{table}[htbp]
\begin{center}
\caption[]{ Characteristics of Prototype \#1 and \#2.}\label{Tab:publ-works}


 \begin{tabular}{clclcl}
  \hline\noalign{\smallskip}
   &  Energy & Prototype \#2 & Prototype \#1 & Prototype \#2 \\
 
  &    & HPD (")  & HPD (") &  EA ($cm^{2}$)           \\
   \hline\noalign{\smallskip}
  & 525 eV & - & - & 33      \\ 
  & 1.49 keV & 111 & 82 & 39      \\ 
  Full flood illumination & 4.51 keV & 106 & - & 38      \\ 
  & 8.04 keV & 99 & - & 23      \\ 
  & 9.9 keV & 93 & - & -      \\ 
\hline\noalign{\smallskip}
 & 1.49 KeV & 109 & 67 & -      \\ 
 Sector A (A')& 8.04 KeV & 96 & 65 & -      \\ 
 & 1.49 KeV & 93 & - & -      \\ 
  \noalign{\smallskip}\hline
  Sector B & 1.49 KeV & 115 & - & -      \\ 
    \noalign{\smallskip}\hline
  Sector C & 1.49 KeV & 115 & - & -      \\ 
   \noalign{\smallskip}\hline
\end{tabular}
\end{center}
\end{table}

The measurement results of Prototype \#1 and \#2 are summarized in Table 2. The systematic error of the HPD is 4" according to the characterization of TRoPIC, while the statistic error is negligible compared with the systematic one.

The prototype has a smaller HPD at higher energy, because it is the figure error that dominated in the PSF test instead of the surface micro-roughness at an energy range of 0.5-10 keV for the tested prototype. In other words, outer layers, corresponding to larger HPD compared with inner layers, contribute less photon at higher energy. For the prototype at 0.5-10 keV, this makes the HPD smaller and this improvement outweighs the degradation on HPD because of X-ray scattering resulted from the surface micro-roughness.

\section{SUMMARY}
\label{sect:analysis}
Imaging X-ray telescopes have been developed at the Institute of Precision Optical Engineering of Tongji University for more than a decade. Currently thermal slumping technology is being used to fabricate mirror substrates. Two X-ray mirror module prototypes assembled at IPOE, based on a conical Wolter-I configuration were tested and calibrated at the PANTER X-ray test facility. For Prototype \#1 with 3 layers, the HPD was determined to be 82" at 1.49 keV. For Prototype \#2 with 21 layers, the comprehensive measurements at several energies were performed to assess the on-axis imaging performance. At our main energy of interest, 1.49 keV (Al-K), measurements of Prototype \#2 give an on-axis HPD of 111" and an effective area of 39 $cm^{2}$. The measurements at the PANTER indicated a reliable prediction of the prototype performance combining the mirror figure evaluation by the LVDT and the simulation by ray-tracing program, which provided us with valuable feedback to help improve the development of our IXTs.

\begin{acknowledgements}
This work was supported under the National Natural Science Foundation of China (No. U1731242, and No. 61621001), and Strategic Priority Research Program of the Chinese Academy of Sciences (No. XDA15010400, No. XDA04060605).
\end{acknowledgements}

\label{lastpage}


\begin{thebibliography}{99}

  \bibitem[Wolter V. H.(1952)]{Wolter V. H.+1952} Wolter V. H. 1952, Ann. Phys. 6(10), 94-114.

  \bibitem[Giacconi R. et al.(1960)]{Giacconi R.+etal+1960} Giacconi R. \& Rossi B. 1960, J. Geophys. Res. 65(2), 773–775.

  \bibitem[VanSpeybroeck L. P. et al.(1972)]{VanSpeybroeck L. P.+etal+1972} VanSpeybroeck L. P. \& Chase R. C. 1972, Appl. Opt. 11(2), 440-445.

  \bibitem[Wolter V. H.(1952)]{Wolter V. H.+1952} Wolter V. H. 1952, Ann. Phys. 445, 286-295.

  
  \bibitem[Werner W. et al.(1977)]{Werner W.+etal+1977} Werner W. 1977, Appl. Opt. 16(3), 764-773.


  \bibitem[Burrows C. J. et al.(1992)]{Burrows C. J.+etal+1992} Burrows C. J., Burg R., Giacconi R. 1992, Astrophys. J. 392(2), 760-765.

  
  \bibitem[Conconi P. et al.(2002)]{Conconi P.+etal+2002} Conconi P., Campana S. 2002, Astron. \& Astrophys. 372(3), 1088-1094.

  
  \bibitem[Thompson P. L. et al.(1999)]{Thompson P. L.+etal+1999} Thompson P. L., Harvey J. E. 1999, Proc. SPIE 3766, 162-172.

  
  \bibitem[Harvey J. E. et al.(2001)]{Harvey J. E.+etal+2001} Harvey J. E., Krywonos A., Thompson P. L., Saha T. T. 2001, Appl. Opt. 40(1), 136-144.

  
  \bibitem[Saha T. T. et al.(2014)]{Saha T. T.+etal+2014} Saha T. T., Zhang W. W., McClelland R. S. 2014, Astron. Telesc. Instrum. 9144, 914418.

  
  \bibitem[Petre R. et al.(1985)]{Petre R.+etal+1985} Petre R., Serlemitsos P. J. 1985, Appl. Opt. 24(12), 1833-1837.

  
  \bibitem[Serlemitsos P. J.(1988)]{Serlemitsos P. J.+1988} Serlemitsos P. J. 1988, Appl. Opt. 27(8), 1447-1452.

  
  \bibitem[Chen S.-H. et al.(2016)]{Chen S.-H.+etal+2016} Chen S.-H., Ma S., Wang Z.-S. 2016, Chin. Opt. Lett. 14(12), 12340.

  
  \bibitem[Liao Y.-Y. et al.(2019)]{Liao Y.-Y.+etal+2019} Liao Y.-Y., Shen Z.-X., Wang Z.-S. 2019, J. Astron. Telesc. Instrum. Syst. 5(1), 014004.

  
  \bibitem[Li T.-P. et al.(2017)]{Li T.-P.+etal+2017} Li T.-P., Xiong S.-L., Zhang S.-N., et al. 2017, Sci. China-Phys. Mech. Astron. 61(3), 031011.

  
  \bibitem[Dong Y.(2014)]{Dong Y.+2014} Dong Y., 2014, Proc. SPIE 9144, 91443O.

  
  \bibitem[Zhang S.-N. et al.(2016)]{Zhang S.-N.+etal+2016} Zhang S.-N., Feroci M., Santangelo A., et al. 2016, Proc SPIE 9905, 99051Q.

  
  \bibitem[Yuan W.-M. et al.(2015)]{Yuan W.-M.+etal+2015} Yuan W.-M., Zhang C., Feng H., et al. 2015, Huazhong Univ. Sci. Tech. J. 23(4), 383.

  
  \bibitem[Wang Z.-S. et al.(2014)]{Wang Z.-S.+etal+2014} Wang Z.-S., Shen Z.-X., Mu B.-Z., et al. 2014, Proc. SPIE 9144, 91441E.

\bibitem[Shen Z.-X. et al.(2018)]{Shen Z.-X.+etal+2018} Shen Z.-X., Yu J., Ma B., et al. 2018, Proc. SPIE 10699, 106991B.


\bibitem[Labov S. E.(1988)]{Labov S. E.+1988} Labov S. E., 1988, Appl. Opt. 27(8), 1465-1469.


\bibitem[Craig W. W. et al.(2011)]{Craig W. W.+etal+2011} Craig W. W., An H., Blaedel K. L., et al, 2011, Proc. SPIE 8147, 81470H.


\bibitem[Koglin J. E. et al.(2004)]{Koglin J. E.+etal+2004} Koglin J. E., Chen C. M. H., Chonko J. C., et al., 2004, Proc. SPIE 5488, 856–867.


\bibitem[Freyberg M. J. et al.(2005)]{Freyberg M. J.+etal+2005} Freyberg M. J., Brauninger H., Burkert W., et al. 2005. Exp. Astron. 20, 405–412.


\bibitem[Freyberg M. J. et al.(2008)]{Freyberg M. J.+etal+2008} Freyberg M. J., Budau B., Burkert W., et al. 2008, Proc. SPIE 7011, 701117.


\bibitem[Burwitz V. et al.(2013)]{Burwitz V.+etal+2013} Burwitz V., Bavdaz M., Pareschi G., et al. 2013, Proc. SPIE 8861, 88611J.


\bibitem[Menz B. et al.(2013)]{Menz B.+etal+2013} Menz B., Brauninger H., Burkert W., et al. 2013, Proc. SPIE 8861, 88611I.

\bibitem[Koglin J. E. et al.(2011)]{Koglin J. E.+etal+2011} Koglin J. E., An H., Barrière N., et al. 2011, Proc. SPIE 8147, 81470J.

\bibitem[Misaki K. et al.(2008)]{Misaki K.+etal+2008} Misaki K., Freyberg M. J., Friedrich P., et al. 2008, Proc. SPIE 7011, 70112Z.

\end{thebibliography}
\end{document}